\pdfoutput=1

\documentclass[11pt]{article}

\usepackage[final]{acl}

\usepackage{times}
\usepackage{latexsym}

\usepackage[T1]{fontenc}

\usepackage[utf8]{inputenc}

\usepackage{microtype}

\usepackage{inconsolata}

\usepackage{multirow}
\usepackage{graphicx}
 \usepackage[normalem]{ulem}
\useunder{\uline}{\ul}{}
\title{ConTReGen:  Context-driven Tree-structured Retrieval for Open-domain Long-form Text Generation}

\author{
 \textbf{Kashob Kumar Roy\textsuperscript{1}} \space
 \textbf{Pritom Saha Akash\textsuperscript{1}} \space
 \textbf{Kevin Chen-Chuan Chang\textsuperscript{1}} \space 
 \textbf{Lucian Popa\textsuperscript{2}}
\\
 \textsuperscript{1}University of Illinois at Urbana-Champaign, USA \space \space 
 \textsuperscript{2}IBM Research, USA
\\
\{kkroy2, pakash2, kcchang\}@illinois.edu, \space \space lpopa@us.ibm.com
}

\begin{document}
\maketitle
\begin{abstract}

Open-domain long-form text generation requires generating coherent, comprehensive responses that address complex queries with both breadth and depth. This task is challenging due to the need to accurately capture diverse facets of input queries. Existing iterative retrieval-augmented generation (RAG) approaches often struggle to delve deeply into each facet of complex queries and integrate knowledge from various sources effectively. This paper introduces ConTReGen, a novel framework that employs a context-driven, tree-structured retrieval approach to enhance the depth and relevance of retrieved content. ConTReGen integrates a hierarchical, top-down in-depth exploration of query facets with a systematic bottom-up synthesis, ensuring comprehensive coverage and coherent integration of multifaceted information. Extensive experiments on multiple datasets, including LFQA and ODSUM, alongside a newly introduced dataset, ODSUM-WikiHow, demonstrate that ConTReGen outperforms existing state-of-the-art RAG models.
\end{abstract}

\section{Introduction}
Large Language Models (LLMs) have transformed various domains through their remarkable performance across a spectrum of tasks. However, LLMs often struggle with generating hallucinated or factually incorrect content, particularly when addressing knowledge-intensive tasks in open-domain settings~\cite{asai2023retrieval,gao2023retrieval}. These limitations typically arise from either the lack of long-tail relevant knowledge or reliance on outdated information embedded within their parameters. To address these challenges, Retrieval-augmented Generation (RAG), therefore, has emerged as a promising solution~\cite{lewis2020retrieval,petroni2020kilt, izacard2023atlas}. RAG enhances LLMs by incorporating external knowledge from a corpus,  effectively reducing hallucinations and factual errors in knowledge-driven tasks such as question answering.

\begin{figure}[h]
  \includegraphics[width=\columnwidth]{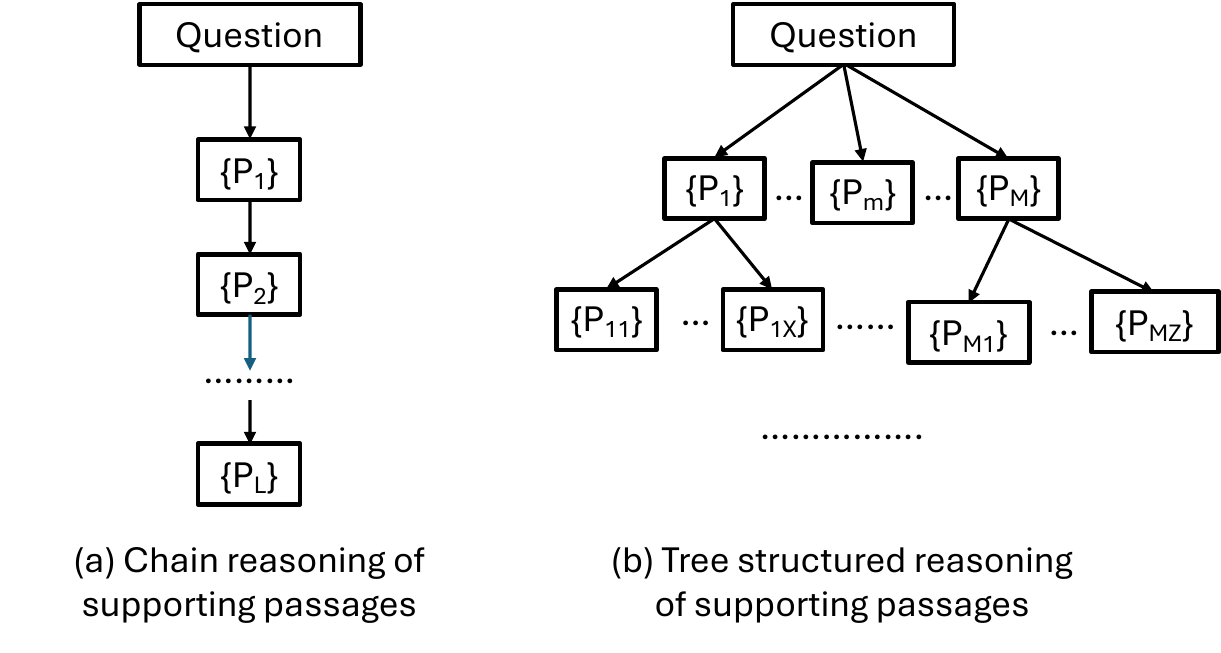}
  \caption{Schematic illustration of retrieval reasoning}
  \label{fig:retrieval_mechanism}
\end{figure}

Recently, significant progress has been made in open-domain question answering, particularly in the realm of short-form answers~\cite{shao2023enhancing, press2022measuring, trivedi2022interleaving, xu2024search}. Rather than adhering to the traditional one-step retrieve-then-generation approach, recent works increasingly adopt iterative retrieval processes~\cite{gao2023retrieval}. These methods utilize either previously generated responses~\cite{shao2023enhancing, trivedi2022interleaving, jiang-etal-2023-active, asai2023self}, or employ a sequence of follow-up questions or interlinked queries~\cite{press2022measuring,xu2024search,qi2019answering}. Such iterative, chain-like strategies have proven particularly effective in short-form factoid question-answering tasks where questions often demand direct, specific pieces of information~\cite{qi2019answering,press2022measuring} and their supporting passages commonly form chain-reasoning of facts~\cite{xiong2020answering, trivedi2022interleaving, xu2021exploiting} as in Figure~\ref{fig:retrieval_mechanism}(a).
\begin{figure}[h]
  \includegraphics[width=\columnwidth]{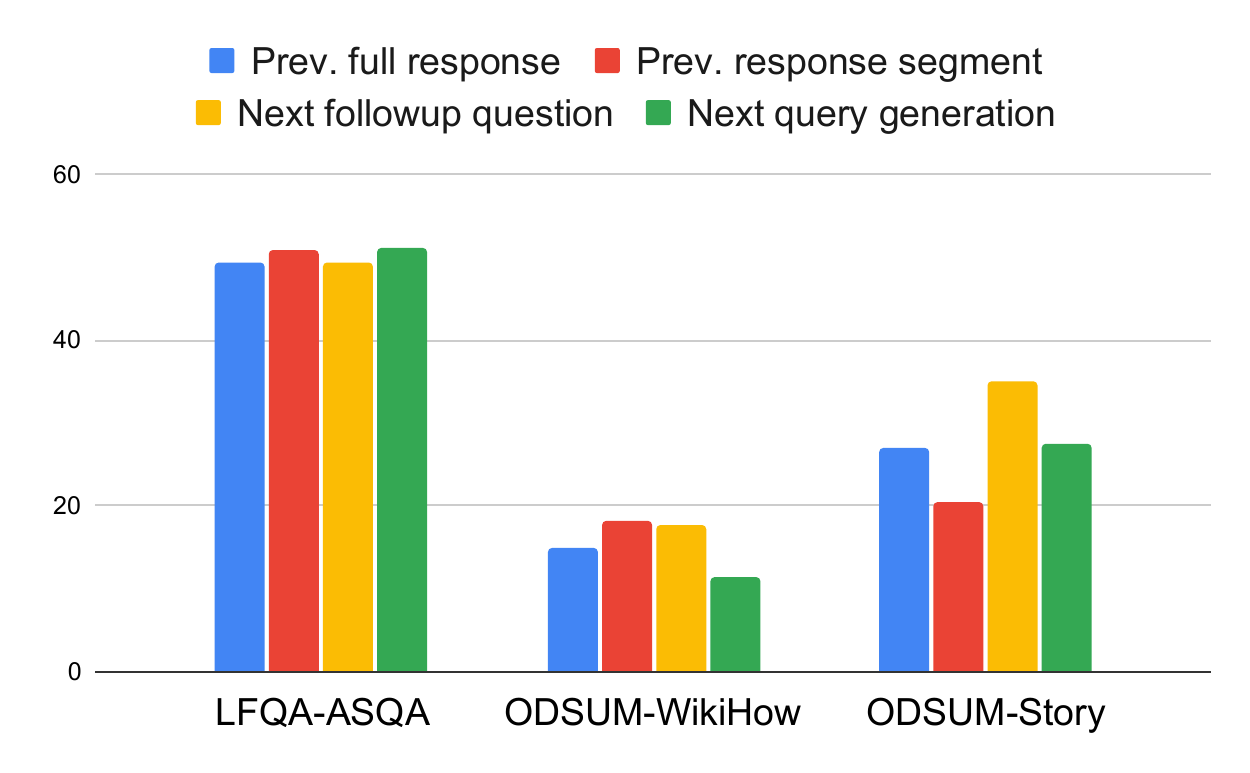}
  \caption{Retrieval Recall Performance. Prev. full response~\cite{shao2023enhancing}, Prev. response segment~\cite{asai2023self}, Next followup question~\cite{press2022measuring,xu2024search}, Next query generation~\cite{khattab2022demonstrate}.}
  \label{fig:initial_retrieval}
\end{figure}

Moving beyond short-form question answering, real-world queries often entail greater complexity and necessitate more comprehensive, detailed responses that encompass multiple facets of the queries. For these multifaceted queries, it is essential to retrieve supporting knowledge from diverse sources, integrating this information into coherent and comprehensive long-form responses. We refer to this scenario as \textit{Open-domain Long-form Text Generation}. This perspective unifies the commonalities between existing long-form question-answering (LFQA)
and open-domain multi-document summarization (ODSUM) tasks, addressing both under a single framework. LFQA~\cite{krishna2021hurdles,fan-etal-2019-eli5} requires not only retrieval of relevant facts from diverse knowledge sources but also integrating them into a coherent paragraph-length answer. Similarly, ODSUM~\cite{giorgi-etal-2023-open} involves the aggregation of information from diverse sources into a unified, coherent summary.

\begin{figure}[h]
  \includegraphics[width=\columnwidth]{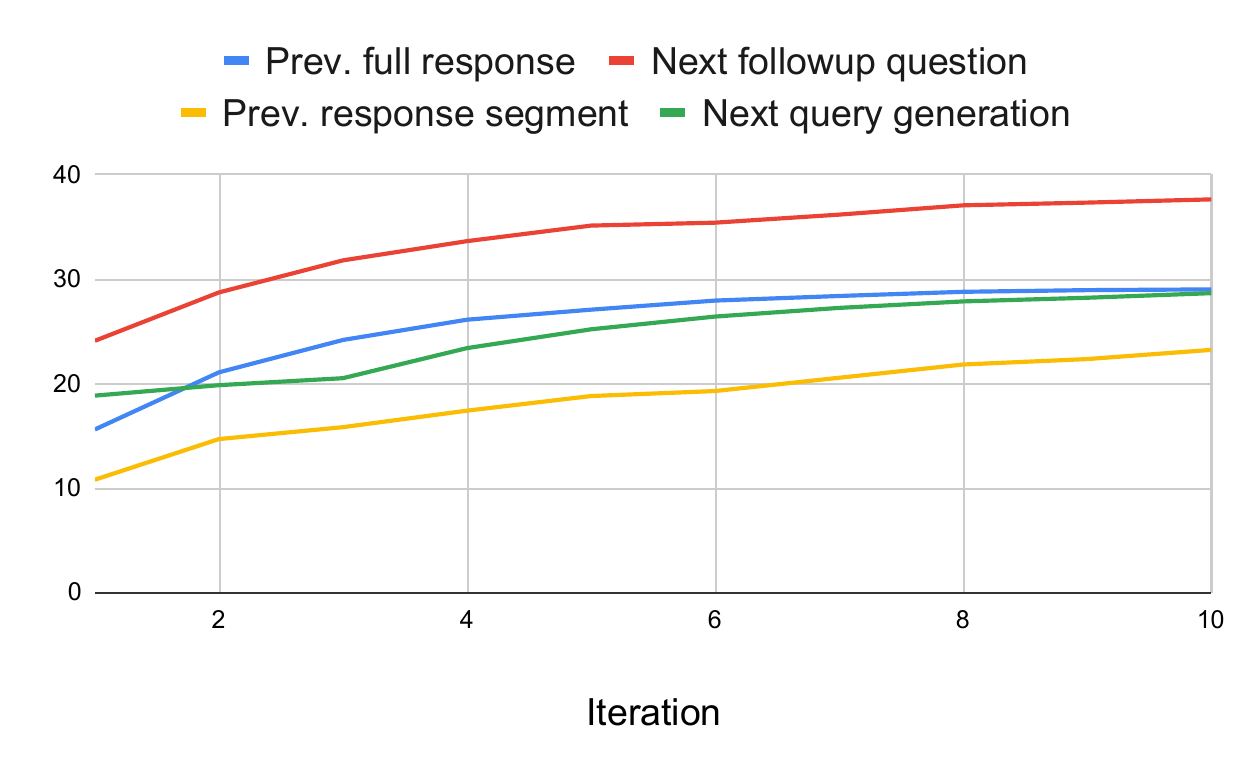}
  \caption{Retrieval Recall Performance per iteration on ODSUM-Story. Prev. full response~\cite{shao2023enhancing}, Prev. response segment~\cite{asai2023self}, Next followup question~\cite{press2022measuring,xu2024search}, Next query generation~\cite{khattab2022demonstrate}}
  \label{fig:retrieverl_per_iterations}
\end{figure}

In our experiments with chain-like iterative approaches on LFQA and ODSUM tasks, we identified several limitations. One notable issue is the low retrieval recall of performance across the datasets in Figure~\ref{fig:initial_retrieval}, especially evident in ODSUM datasets where the need for multi-faceted information is more implicit. Secondly, we observed that these iterative approaches quickly reach a plateau in terms of retrieval recall as demonstrated in Figure~\ref{fig:retrieverl_per_iterations}. In other words, they fail to retrieve new relevant passages during subsequent iterations. 

One of the key reasons behind these limitations is the simplistic modeling of the reasoning of retrieving passages as a chain of semantic aspects as shown in Figure~\ref{fig:retrieval_mechanism}(a). However, when the retrieval of information spans multiple facets and originates from diverse sources, it naturally forms a complex retrieval reasoning structure as illustrated in Figure~\ref{fig:retrieval_mechanism}(b). For example, a query like \textit{'How to Detect Hidden Cameras and Microphones'} may require information from multiple facets such as \textit{'Conducting a Physical Search'}, \textit{'Searching for Electrical Signals'}, and so on, where each facet may need information from its multiple subfacets. For instance, this \textit{'Conducting a Physical Search'} facet needs information about its subfacets such as \textit{'Investigating your smoke detectors and other electronics.'},  \textit{'Using to check for two-way mirrors'}, etc. Consequently, this yields a tree structure of facets that organizes the need for information from the broadest to the most specific ones.

Furthermore, the method of formulating subsequent retrieval queries plays a crucial role in thoroughly exploring all facets of an input query. Current iterative approaches often generate the next query based on segments of the previous response or all previously retrieved information. We argue that this sequential approach to query formulation inherently restricts the retrieval search scope. This is because it tends to allow a few aspects of the information to dominate the retrieval process, potentially leading to a biased exploration where certain facets are disproportionately emphasized, while others may remain underexplored or unexplored. 


\begin{figure*}[]
  \includegraphics[width=0.96\textwidth]{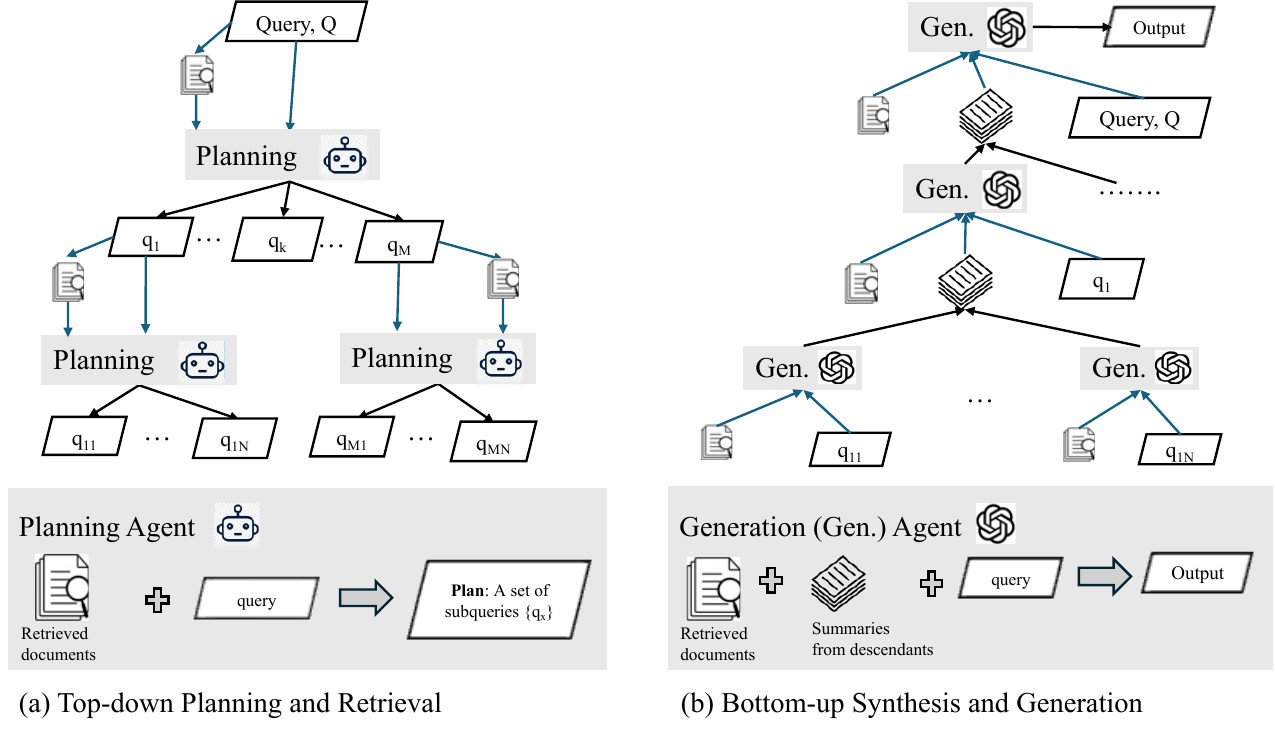}
  \caption {ConTReGen Framework.}
  \label{fig:framework}
\end{figure*}

To accommodate these findings, we introduce a novel tree-structured retrieval augmented generation framework (ConTReGen) that conceptualizes the retrieval process as a hierarchical exploration of an input question or query’s various facets, organized in a tree structure. Each branch of this tree represents a specific facet, enabling a systematic and comprehensive exploration of the query. This method enables an in-depth exploration of each query facet, from the broadest facet down to the finer specific through a top-down strategy. Additionally, to facilitate a more thorough aggregation and synthesis of information, we leverage a bottom-up generation technique that synthesizes information from the leaf nodes upwards, ensuring that all retrieved data contributes cohesively to the final output. This integrated approach significantly enhances the depth and relevance of the retrieved content, contributing to more coherent and contextually rich text generation.

To evaluate the efficacy of our framework on open-domain long-form text generation, we conducted experiments using both the LFQA and ODSUM datasets, including ASQA~\cite{stelmakh-etal-2022-asqa} and ODSUM-Story~\cite{zhou2023odsum}. Additionally, we introduced a new, large-scale open-domain summarization dataset named ODSUM-WikiHow. The experimental results demonstrate that ConTReGen significantly outperforms state-of-the-art RAG baselines across all three datasets.

\section{Methodology}

Our framework addresses the task of context-driven Long-Form Text Generation which aims to generate a comprehensive long text, $\mathbf{Y}$ for an input query or question, denoted as, $\mathbf{Q}$ by leveraging relevant passages from a given corpus $\mathbf{C}$. 

Our proposed framework is structured into a two-stage process designed to enhance the retrieval and generation of long-form text responses: 1) during the top-down planning and retrieval stage as shown in Figure~\ref{fig:framework}(a), the process begins by retrieving passages directly related to the input query or question. Utilizing the insights gained from these initial passages, the planning agent generates a series of subquestions or subqueries, which serve as a plan for deeper exploration. Each of these subquestions is recursively used to generate successive plans to continually expand the search scope of relevant passages. This recursive planning and retrieval results in a comprehensive tree structure, where each node represents a query and its associated retrieved passages; 2) in Figure~\ref{fig:framework}(b), the bottom-up synthesis and generation stage of the framework begins at the leaf nodes, summarizes the relevant information from retrieved documents. The model then works its way upwards, synthesizing the retrieved documents and the summarized information from descendant nodes to generate text responses for each intermediate node. This ascending integration continues until it reaches the root of the tree, producing a final response that comprehensively addresses the original input question or queries. This methodical approach ensures that every facet of the query is thoroughly explored, allowing for a detailed and well-structured response generation. 

\subsection{Top-down Planning and Retrieval}
The concept of top-down planning and retrieval stems from observing how humans approach complex information-seeking tasks~\cite{kuhlthau1991inside, marchionini1995information}. When faced with a multifaceted query, individuals naturally decompose the complex query into smaller, specific facet-focused manageable sub-queries~\cite{wei2013survey}. Our framework aims to mimic this human-like approach to information retrieval by structuring the process as a hierarchical exploration of various facets of a complex query or question. This top-down planning and retrieval stage is designed to systematically break down the query into subquestions/subqueries from multiple facets that guide the retrieval process towards a more thorough and nuanced aggregation of relevant information. 
\paragraph{Planning.} The process begins with analyzing the input query or question and its initial retrieved passages. We employ LLM as the planning agent in Figure~\ref{fig:framework}(a) to generate a series of sub-questions as a plan, ensuring a broad and diverse exploration of the query or question. Each sub-question targets a specific aspect/facet of the query.  
 
\begin{figure}[!htbp]
  \includegraphics[width=0.96\columnwidth]{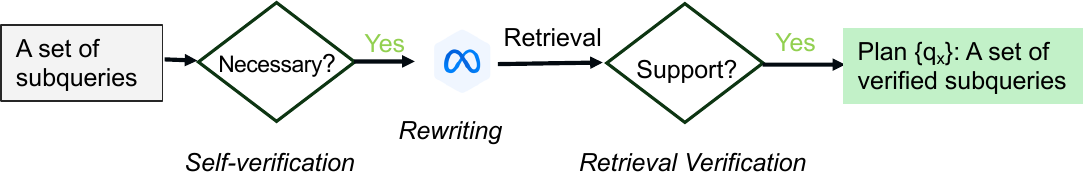}
  \caption {Two-step Verification.}
  \label{fig:verification}
\end{figure}

\paragraph{Verification.} Ideally, each sub-question should be essential for addressing the original question or query and should be contextually well-informed to ensure it can retrieve passages relevant to the original query. We apply a two-step verification process as shown in Figure~\ref{fig:verification}: i) we leverage LLM to predict whether a sub-question is required to address the main query/question. If the sub-question is identified as necessary, the LLM is used to rewrite it as a self-sufficient and contextually-rich search query. This step ensures that each sub-question is both relevant and comprehensive, making it capable of independently guiding the retrieval process;
ii) Each sub-question is then used individually as a query to retrieve passages. We evaluate whether these retrieved passages are relevant to the original query. For this relevancy assessment, we again use LLMs to ensure that the retrieved passages effectively contribute to addressing the original query. This verification process ensures that the decomposition of the original query into sub-questions is not only comprehensive and contextually accurate but also effective in retrieving relevant information. 

\paragraph{Recursive Exploration.} Once obtained verified set of subquestions for the original input, we recursively repeat the same planning-then-verification strategy to generate successive plans for each subquestion, continuously expanding the retrieval scope. This recursive process ensures an in-depth exploration of each query dimension, from the broadest facets to the finer ones. It continues until no further plans are needed or the predefined maximum depth is reached. This stage results in a detailed and well-structured tree where each node represents a specific sub-question and its corresponding relevant passages and the root node corresponds to the original query.

\subsection{Bottom-up Synthesis and Generation}
This stage, as shown in Figure~\ref{fig:framework}(b), focuses on synthesizing the retrieved information to produce a coherent and comprehensive long-form response that addresses the original query. Key advantages of this bottom-up strategy include: i) utilizing a large volume of retrieved passages by summarizing information at each descendant level without exceeding the input token length limit, and ii) effectively filtering out irrelevant or less important details, which enhances the quality of the final response. The process begins at the leaf nodes of the tree, which contain the most specific subquestions and their retrieved passages. Each leaf subquestion represents a fine-grained aspect of the original query. We employ LLM to summarize the key information from the retrieved passages that are relevant to the sub-question.
answer to complex, open-domain questions.
\begin{table}[]
\centering
\large
\resizebox{\columnwidth}{!}{%
\begin{tabular}{ccc}
\hline
Dataset                  & ODSUM-Story & ODSUM-WikiHow \\ \hline
Corpus Size              & 1138        & 506295        \\
Number of Queries        & 635         & 25896         \\
Avg \# docs per Query    & 8.78        & 15.26        \\
Avg Doc Length     & 621.93      & 81.29         \\
Avg Query Length   & 10.18       & 6.83          \\
Avg Summary Length & 274.65      & 124.24        \\ \hline
\end{tabular}%
}
\caption{Statistics of ODSUM-Story and ODSUM-WikiHow: Avg length in terms of words.}
\label{tab:odsum_dataset_comparison}
\end{table}

From the leaf nodes, the synthesis and generation process works its way upwards through the tree. Similar to the leaf node, each intermediate node contains a subquestion and its retrieved passages. Each intermediate node integrates the summarized information from its child nodes with its retrieved passages. This involves merging the key insights and details from the descendant queries to form a coherent and contextually rich response for each sub-question represented by the node. This upward process continues through each level of the tree. At each step, the synthesized response from a lower level is combined with the retrieved passages at the current level. The process stops at the root node, which represents the original query or question. At this point, the final comprehensive response is generated to address the original query, effectively utilizing the information gathered throughout the process. 
\section{WikiHow: A new ODSUM dataset}
Open-domain summarization (ODSUM) is to generate a comprehensive summary for a given query by gathering information from a large index of documents. One of the key challenges in open-domain summarization (ODSUM) is the scarcity of datasets that encompass real-world queries alongside a well-curated document corpus~\cite{giorgi-etal-2023-open}. While a recent work~\cite{zhou2023odsum} constructed a dataset ODSUM-Story from SQuALITY~\cite{wang-etal-2022-squality}, question-based abstraction summarization for short stories. However, the corpus size and the diversity of queries (only from the story domain) are limited as shown in Table~\ref{tab:odsum_dataset_comparison}. To address this gap, we introduce a large-scale open-domain summarization derived from WikiHow\footnote{https://www.wikihow.com/Main-Page} articles. Existing datasets sourced from WikiHow, such as ~\cite{cohen2021wikisum,koupaee2018wikihow,boni2021howsumm}, are constructed for traditional text summarization tasks. Moreover, A recent WikiHowQA dataset~\cite{bolotova2023wikihowqa} focuses on multi-document long-form question answering with supporting documents sourced from article reference links.  Unfortunately, its corpus of documents is not publicly available, which limits its utility in LFQA.  In addition, it utilizes only 11,746 queries from a potential 25,896. On the other hand, we have developed the ODSUM-WikiHow dataset, which includes 25,896 queries, each paired with human-written, coherent summaries and supported by a substantial corpus of 506,295 documents.
\paragraph{Dataset Construction} We process the article titles as input queries and use author-provided summaries as the ground-truth summaries. WikiHow articles are categorized into two types: one type describes tasks using a single method detailed in successive steps, while the other type outlines multiple methods, with each method comprising several steps. To prepare the corpus, we exploit this structured format where each step is discussed in a separate paragraph. We treat each paragraph as an individual document, which makes this dataset particularly retrieval-intensive. As a result, on average, each query is supported by 15 documents, enhancing the complexity and depth of the retrieval challenge. The detailed stats are reported in Table~\ref{tab:odsum_dataset_comparison}.
\begin{table*}[]
\centering
\resizebox{\textwidth}{!}{%
\begin{tabular}{|c|ccccccc|ccccccc|}
\hline
\multirow{3}{*}{} & \multicolumn{7}{c|}{ODSUM-Story}                                                                                                                                                                                                                             & \multicolumn{7}{c|}{ODSUM-WikiHow}                                                                                                                                                                                                                         \\ \cline{2-15} 
                  & \multicolumn{1}{c|}{\multirow{2}{*}{RL}} & \multicolumn{1}{c|}{\multirow{2}{*}{BS}} & \multicolumn{2}{c|}{NLI}                                              & \multicolumn{3}{c|}{UniEval}                                                               & \multicolumn{1}{c|}{\multirow{2}{*}{RL}} & \multicolumn{1}{c|}{\multirow{2}{*}{BS}} & \multicolumn{2}{c|}{NLI}                                            & \multicolumn{3}{c|}{UniEval}                                                               \\ \cline{4-8} \cline{11-15} 
                  & \multicolumn{1}{c|}{}                    & \multicolumn{1}{c|}{}                    & \multicolumn{1}{c|}{Ent.}           & \multicolumn{1}{c|}{Con.}           & \multicolumn{1}{c|}{Rel.}           & \multicolumn{1}{c|}{Coh.}           & Cons.          & \multicolumn{1}{c|}{}                    & \multicolumn{1}{c|}{}                    & \multicolumn{1}{c|}{Ent.}          & \multicolumn{1}{c|}{Con.}          & \multicolumn{1}{c|}{Rel.}           & \multicolumn{1}{c|}{Coh.}           & Cons.          \\ \hline
RetGen            & \multicolumn{1}{c|}{15.19}               & \multicolumn{1}{c|}{48.19}               & \multicolumn{1}{c|}{10.75}          & \multicolumn{1}{c|}{44.52}          & \multicolumn{1}{c|}{64.02}          & \multicolumn{1}{c|}{63.41}          & 64.91          & \multicolumn{1}{c|}{27.51}               & \multicolumn{1}{c|}{60.17}               & \multicolumn{1}{c|}{18.64}         & \multicolumn{1}{c|}{11.67}          & \multicolumn{1}{c|}{81.71}          & \multicolumn{1}{c|}{81.11}          & 80.55          \\
IterRetGen        & \multicolumn{1}{c|}{14.11}               & \multicolumn{1}{c|}{49.47}               & \multicolumn{1}{c|}{13.8}           & \multicolumn{1}{c|}{38.95}          & \multicolumn{1}{c|}{69.15}          & \multicolumn{1}{c|}{70.17}          & 71.01          & \multicolumn{1}{c|}{24.32}               & \multicolumn{1}{c|}{59.28}               & \multicolumn{1}{c|}{18.24}         & \multicolumn{1}{c|}{10.76}         & \multicolumn{1}{c|}{79.41}          & \multicolumn{1}{c|}{78.41}          & 78.24          \\
Self-Ask          & \multicolumn{1}{c|}{17.29}               & \multicolumn{1}{c|}{50.88}               & \multicolumn{1}{c|}{12.99}          & \multicolumn{1}{c|}{24.68}          & \multicolumn{1}{c|}{85.5}           & \multicolumn{1}{c|}{85.64}          & 86.23          & \multicolumn{1}{c|}{28.99}               & \multicolumn{1}{c|}{60.66}               & \multicolumn{1}{c|}{{\ul 22.79}}   & \multicolumn{1}{c|}{7.98}          & \multicolumn{1}{c|}{82.97}          & \multicolumn{1}{c|}{82.13}          & 82.97          \\
SearChain         & \multicolumn{1}{c|}{12.9}                & \multicolumn{1}{c|}{46.23}               & \multicolumn{1}{c|}{4.41}           & \multicolumn{1}{c|}{51.47}          & \multicolumn{1}{c|}{76.79}          & \multicolumn{1}{c|}{72.65}          & 75.14          & \multicolumn{1}{c|}{26.52}               & \multicolumn{1}{c|}{55.59}               & \multicolumn{1}{c|}{14.25}         & \multicolumn{1}{c|}{23.68}         & \multicolumn{1}{c|}{82.01}          & \multicolumn{1}{c|}{80.5}           & 80.94          \\
DSP               & \multicolumn{1}{c|}{17.09}               & \multicolumn{1}{c|}{{\ul 51.52}}         & \multicolumn{1}{c|}{{\ul 14.15}}    & \multicolumn{1}{c|}{{\ul 16.34}}    & \multicolumn{1}{c|}{{\ul 88.2}}     & \multicolumn{1}{c|}{{\ul 87.9}}     & {\ul 89}       & \multicolumn{1}{c|}{25.22}               & \multicolumn{1}{c|}{58.21}               & \multicolumn{1}{c|}{17.3}          & \multicolumn{1}{c|}{11.5}          & \multicolumn{1}{c|}{{\ul 85.15}}    & \multicolumn{1}{c|}{{\ul 85.37}}    & {\ul 86.28}    \\
Self-RAG          & \multicolumn{1}{c|}{{\ul 17.81}}         & \multicolumn{1}{c|}{50.16}               & \multicolumn{1}{c|}{11.73}          & \multicolumn{1}{c|}{\textbf{16.26}} & \multicolumn{1}{c|}{55.81}          & \multicolumn{1}{c|}{52.6}           & 54.81          & \multicolumn{1}{c|}{19.77}               & \multicolumn{1}{c|}{56.24}               & \multicolumn{1}{c|}{21.35}         & \multicolumn{1}{c|}{\textbf{7.76}} & \multicolumn{1}{c|}{83.95}          & \multicolumn{1}{c|}{84.07}          & 83.78          \\

ConTReGen         & \multicolumn{1}{c|}{\textbf{19.33}}      & \multicolumn{1}{c|}{\textbf{54.01}}      & \multicolumn{1}{c|}{\textbf{21.35}} & \multicolumn{1}{c|}{{16.98}} & \multicolumn{1}{c|}{\textbf{89.28}} & \multicolumn{1}{c|}{\textbf{89.27}} & \textbf{89.72} & \multicolumn{1}{c|}{\textbf{34.88}}      & \multicolumn{1}{c|}{\textbf{62.21}}      & \multicolumn{1}{c|}{\textbf{24.3}} & \multicolumn{1}{c|}{{\ul 7.89}}    & \multicolumn{1}{c|}{\textbf{86.83}} & \multicolumn{1}{c|}{\textbf{86.76}} & \textbf{87.19} \\ \hline
\end{tabular}%
}
\caption{ODSUM Performance Comparison}
\label{tab:odsum_performance_comparison}
\end{table*}
\section{Experimental Setup}
\subsection{Datasets and Evaluation Metrics}
\paragraph{Open-domain Summarization}
 In addition to ODSUM-WikiHow, we use the recent ODSUM-Story~\cite{zhou2023odsum} dataset in our experiments. In this dataset, inputs are queries and a corpus of documents, the task is to generate a summary by retrieving multiple relevant documents from the corpus. For evaluation, we use commonly used Rouge-L (R-L), BertScore (BS), NLI-based Entailment (Ent.) and Contradiction (Con.) scores~\cite{liu2019roberta}. Additionally, we use UNIEVAL~\cite{zhong2022towards} to assess the coherence, consistency, and relevance of the generated summaries.


\paragraph{Long-form QA}
In this work, we use the ASQA ~\cite{stelmakh-etal-2022-asqa} dataset where inputs are ambiguous questions with multiple interpretations, and outputs should cover correct answers for all of them. To evaluate the performance on this dataset, we use a set of questions that requires 5 or more evidence passages. disambiguation metrics defined as a good long-form answer to an ambiguous question should contain short answers to all disambiguated questions as well as the context necessary to understand the source of ambiguity and the relationship between the short answers in~\cite{stelmakh-etal-2022-asqa}. EM (string Exact Match) is the fraction of disambiguations for which the corresponding short answer is present in the long answer. DA-F1 (Disambig-F1) is the fraction of disambiguated questions that can be answered from the predicted long answers. DR is the geometric mean of DA-F1 and ROUGE-L.
\subsection{Baselines}
\textbf{RetGen} retrieves passages a single time using the input query itself and then utilizes these passages to generate the response. \\
\textbf{IterRetGen}~\cite{shao2023enhancing} iteratively retrieves passages by using the previously generated response as the next query to retrieve passages in subsequent iterations and use them to update and refine the previously generated response. \\ 
\textbf{Self-Ask}~\cite{press2022measuring}  utilizes an elicitive prompting technique that involves generating follow-up questions based on the previous retrieved knowledge. These follow-up questions are then used to fetch additional relevant passages and use them to generate intermediate responses. \\
\textbf{SearChain}~\cite{xu2024search} utilizes a structured, iterative interaction between LLMs and retriever to enhance reasoning in complex tasks. It dynamically generates and refines a chain of query-answer pairs that are verified through IR. Unlike our ConTReGen approach, which focuses on constructing tree-structured reasoning by exploring each query facet in-depth, it refines the generated CoQ at each round. However, the linear chain-of-query approach often fails to fully explore all relevant facets of a query, leading to potential gaps in the information retrieved.\\
\textbf{DSP}~\cite{khattab2022demonstrate} methodically handles complex queries by first demonstrating the desired query processing behavior, then searching relevant information through iterative decompositions of the query into simpler sub-queries, and finally generating the response by synthesizing all information.\\
\textbf{Self-RAG}~\cite{asai2023self} leverages adaptive retrieval and self-reflection into the generation process. This approach allows the LLM to dynamically retrieve information when needed and use reflection tokens to assess and improve the relevance and factual accuracy of its outputs.
\begin{table}[]
\centering
\resizebox{\columnwidth}{!}{%
\begin{tabular}{|c|c|c|c|}
\hline
           & ODSUM-Story    & WikiHow-ODSUM  & LFQA-ASQA           \\ \hline
RetGen     & 16.14          & 13.99          & 36.56          \\
IterRetGen & 26.98          & 18.01          & 49.42          \\
Self-Ask   & {\ul 36.79}    & {\ul 22.3}     & 51.5           \\
DSP        & 27.61          & 14.94          & {\ul 53.39}    \\
Self-RAG   & 21.14          & 19.37          & 46.27          \\
ConTReGen  & \textbf{57.69} & \textbf{42.44} & \textbf{57.38} \\ \hline
\end{tabular}%
}
\caption{Retrieval Recall Performance}
\label{tab:retrieval_performance}
\end{table}

\subsection{Implementation Details}
In our experiments, we utilize LLAMA3-8B~\cite{llama3modelcard} as the Large Language Model (LLM) for ConTReGen and all baselines, except for Self-RAG. For Self-RAG, which requires fine-tuning of the LLM with additional reflection tokens, we use the provided trained LLAMA2-7B model. Across all methods, we use the pre-trained Contriever~\cite{gautier2022unsupervised} as the dense retriever, setting the number of retrieved passages to topk = 5 for ASQA and ODSUM-WikiHow datasets, and topk = 3 for ODSUM-Story. For iterative approaches, the maximum number of iterations is set to \{5, 10\}, while for ConTReGen, the maximum depth of the tree is set to 2. These settings are used as the default unless explicitly mentioned otherwise. We have used 1-shot prompting in both modules.

\section{Experimental Results}
\subsection{Retrieval Performance}
Table~\ref{tab:retrieval_performance} presents a comparison of the retrieval recall performance between our proposed ConTReGen approach and existing baseline methods. RetGen performs a single round of context passage retrieval from the corpus, while the other baseline methods utilize iterative retrieval techniques. For these baselines, we evaluate performance after five iterations. The results demonstrate that ConTReGen significantly outperforms all baseline approaches. Notably, ConTReGen achieves substantial improvements in recall across various datasets: it shows a 20.9 points increase on ODSUM-Story, a 20.14 points increase on WikiHow-ODSUM, and a 3.99 points increase on ASQA, compared to the second-best methods.
\subsection{ODSUM Performance}
Table~\ref{tab:odsum_performance_comparison} illustrates the performance of ConTReGen compared to several baseline methods on open-domain summarization tasks: ODSUM-Story and ODSUM-WikiHow datasets. Our approach, ConTReGen, consistently outperforms other methods across multiple metrics, demonstrating its effectiveness in generating high-quality summaries.

ConTReGen achieves the highest scores in most metrics. It records a Rouge-L (RL) score of 19.33,  BertScore (BS) of 54.01 on the ODSUM-Story dataset, and On ODSUM-WikiHow, RL score of 34.88, BS of 62.21 surpassing the second-best methods by 1.52, 2.49, 5.89 and 1.55 points, respectively. This improvement highlights ConTReGen’s ability to accurately capture and reproduce more relevant information.

Going beyond token-level similarity-based metrics, we use NLI-based Entailment and Contradict scores to evaluate how generated summary sentences are logically aligned to the reference summary. ConTReGen excels with an entailment (Ent.) score of 21.35 on ODSUM-Story and 24.3 on ODSUM-WikiHow, outperforming all other baselines. It achieves competitive contradiction scores to Self-RAG. As Self-RAG utilizes the fine-tuned LLMs with reflection tokens to not only filter out the relevant passages but also to select best-generated text segment, it helps to improve generation quality with less contradictory information. 

Furthermore, we use pre-trained QA-based UNIEVAL~\cite{zhong2022towards} metric to evaluate the generated summary on multiple dimensions: Relevance (Rel.),  Coherence (Coh.), and Consistency (Cons.). On both datasets, ConTReGen shows superior performance across all dimensions, indicating its effectiveness not only in retrieving relevant information but also in integrating it cohesively and consistently into the generated summaries.
\begin{table}[]
\centering
\resizebox{\columnwidth}{!}{%
\begin{tabular}{|c|r|c|c|cc|}
\hline
\multirow{2}{*}{} & \multicolumn{1}{c|}{\multirow{2}{*}{EM}} & \multirow{2}{*}{DA-F1} & \multirow{2}{*}{DR} & \multicolumn{2}{c|}{NLI}                             \\ \cline{5-6} 
                  & \multicolumn{1}{c|}{}                    &                        &                     & \multicolumn{1}{c|}{Ent.}           & Con.           \\ \hline
RetGen            & 28.91                                    & 20.64                  & 18.95               & \multicolumn{1}{c|}{24.58}          & 29.27          \\
IterRetGen        & 35.52                                    & 24.56                  & 25.82               & \multicolumn{1}{c|}{\ul33.82}          & {\ul 19.85}    \\
Self-Ask          & {\ul37.02}                                    & {\ul 25.08}            & 26.89               & \multicolumn{1}{c|}{31.99}          & 22.22          \\
SearChain         & 28.26                                    & 13.47                  & 18.99               & \multicolumn{1}{c|}{7.79}           & 53.89          \\
DSP               & 36.22                                    & 24.59                  & {\ul 27.03}         & \multicolumn{1}{c|}{29.91}          & \textbf{16.17} \\
Self-RAG          & 22.41                                    & 12.09                  & 18.27               & \multicolumn{1}{c|}{21.63}          & 38.33          \\
ConTReGen         & \textbf{41.16}                           & \textbf{30.23}         & \textbf{30.31}      & \multicolumn{1}{c|}{\textbf{41.51}} & {21.4}     \\ \hline
\end{tabular}%
}
\caption{LFQA Performance on ASQA dataset}
\label{tab:lfqa_performance}
\end{table}
\subsection{LFQA Performance}
As shown in Table~\ref{tab:lfqa_performance}, ConTReGen achieves the highest Exact Match (EM) score of 41.16 and DA-F1 score of 30.23, indicating its superior ability to recall short answers for more disambiguated questions. Furthermore, the larger DR score of 30.31 highlights the effectiveness of ConTReGen in terms of factual accuracy and text quality as this score combines both Rouge-L and DA-F1 scores. Additionally, ConTReGen achieves NLI entailment (Ent.) scores of 41.51, outperforming baselines by a large margin. However, DSP gets lower contradiction scores (Con.) than ConTReGen, because of DSP's advanced Chain-of-prompting compared to ConTReGen's one-shot prompting.
\subsection{Ablation Study}
\textbf{Impact of Retrieval enhancements:} To further analyze the contribution of retrieval in downstream tasks, we conducted an experiment on the ODSUM-Story dataset where we replaced our Bottom-up Synthesis and Generation (BSG) module with a simple generation module. This simple generation module takes the top K=3 passages retrieved by our retrieval system and generates the response once. We compared this performance with RetGen, which directly retrieves the top K=3 passages from the corpus index and generates a response. Table~\ref{tab:ablation_study} shows that ConTRe w/o BSG outperforms the RetGen approach. \\
\textbf{Impact of Bottom-up Synthesis and Generation}: The structure of our retrieval systems encourages to leverage the BSG module in the generation process. In Table~\ref{tab:ablation_study}, we can see that BSG module improves the overall performance across all metrics.
\begin{table}[]
\centering
\resizebox{\columnwidth}{!}{%
\begin{tabular}{|c|c|c|c|}
\hline
               & R-L            & BERTScore      & NLI Entailment \\ \hline
RetGen         & 15.19          & 48.19          & 10.75          \\ 
ConTRe w/o BSG & 18.62          & 53.45          & 19.23          \\
ConTReGen      & \textbf{19.33} & \textbf{54.01} & \textbf{21.35} \\ \hline
\end{tabular}%
}
\caption{Ablation Study}
\label{tab:ablation_study}
\end{table}
\subsection{Analysis}
\begin{table}[]
\centering
\resizebox{\columnwidth}{!}{%
\begin{tabular}{|c|cc|cc|cc|}
\hline
                                   & \multicolumn{2}{c|}{ODSUM-Story}                & \multicolumn{2}{c|}{ODSUM-WikiHow}              & \multicolumn{2}{c|}{LFQA-ASQA}                       \\ \cline{2-7} 
\multirow{-2}{*}{}                 & \multicolumn{1}{c|}{ReP} & NReP       & \multicolumn{1}{c|}{ReP} & NReP       & \multicolumn{1}{c|}{ReP} & NReP       \\ \hline
IterRetGen & \multicolumn{1}{c|}{53.8}      & 8.46           & \multicolumn{1}{c|}{41.43}     & 6.18           & \multicolumn{1}{c|}{72.35}     & 16.96          \\ 
Self-Ask   & \multicolumn{1}{c|}{61.47}     & {21.74}          & \multicolumn{1}{c|}{50.11}     & 8.28           & \multicolumn{1}{c|}{77.49}     & 14.72          \\ 
DSP        & \multicolumn{1}{c|}{47.49}     & 13.8           & \multicolumn{1}{c|}{36.3}      & 4.17           & \multicolumn{1}{c|}{76.36}     & 13.69          \\ 
Self-RAG   & \multicolumn{1}{c|}{40.02}     & 8.01           & \multicolumn{1}{c|}{47.63}     & 9.25           & \multicolumn{1}{c|}{72.75}     & 8.86           \\ 
ConTReGen  & \multicolumn{1}{c|}{\textbf{81.11}}     & \textbf{45.62} & \multicolumn{1}{c|}{\textbf{76.67}}     & \textbf{25.18} & \multicolumn{1}{c|}{\textbf{82.46}}     & \textbf{21.88} \\ \hline
\end{tabular}%
}
\caption{Retrieval Reasoning Analysis}
\label{tab:retrieval_dissection}
\end{table}
\subsubsection{Retrieval Reasoning Analysis}
We analyze the retrieval reasoning of iterative approaches and ConTReGen by examining the structural retrieval relationships among evidence passages. For each query and its associated evidence passages, we construct a relationship graph. In this graph, a directed edge from passage A to passage B (A$\rightarrow$B) indicates that passage A can retrieve passage B from the corpus. The query itself acts as the root node, with edges leading from the query to an evidence passage (Q$\rightarrow$A) if the query can directly retrieve Passage A. Based on this graph structure, we divide evidence passages into two categories: i) Reachable Passages (ReP), which are reachable from the root query through the graph, and ii) Non-Reachable Passages (NReP), which are not reachable. We assess and compare the recall performance for each category as detailed in Table~\ref{tab:retrieval_dissection}.
This analysis shows that ReP recall scores are significantly higher than those for NReP, illustrating that iterative approaches excel when there is direct retrieval connectivity among passages. However, these methods often fail to capture passages that lack explicit retrieval connectivity. Conversely, ConTReGen shows superior capability of retrieving both types of passage. This is because hierarchical planning not only ensures the effective utilization of retrieved knowledge to exploit explicit retrieval connectivity among passages but also establishes new implicit connections of passages. 
\begin{table}[]
\centering
\resizebox{\columnwidth}{!}{
\begin{tabular}{|c|c|c|c|c|c|}
\hline
RetGen                       & IterRetGen & Self-Ask & DSP  & Self-RAG & ConTReGen \\ \hline
46.2 & 52.87      & {\ul61.31}    & 43.4 & 51.6     & \textbf{80.65}     \\ \hline
\end{tabular}
}
\caption{Facet Coverage Analysis}
\label{tab:aspect_coverage}
\end{table}

\subsubsection{Facet Coverage Analysis}
To assess the facet coverage of input queries by each approach, we analyze a set of queries from the ODSUM-WikiHow dataset that require information from multiple methods, treating each method as a distinct facet. We considered a facet as covered if an approach successfully retrieves at least one passage corresponding to that facet. According to the results presented in Table~\ref{tab:aspect_coverage}, iterative approaches generally manage to retrieve passages from facets directly accessible by the query itself. In contrast, ConTReGen not only retrieves passages from aspects directly accessible by the query but also effectively covers additional aspects not directly retrievable by the query.
\begin{table}[]
\centering
\resizebox{\columnwidth}{!}{%
\begin{tabular}{|c|c|c|c|cc|}
\hline
\multirow{2}{*}{}                                             & \multirow{2}{*}{EM} & \multirow{2}{*}{DA-F1} & \multirow{2}{*}{DR} & \multicolumn{2}{c|}{NLI}                             \\ \cline{5-6} 
                                                              &                     &                        &                     & \multicolumn{1}{c|}{Ent.}           & Con.           \\ \hline
DSP                                                           & 36.22               & 24.59                  & { 27.03}         & \multicolumn{1}{c|}{29.91}          & {16.17} \\ 
\begin{tabular}[c]{@{}c@{}}DSP w/ ConTRe\end{tabular}      & \textbf{50.34}               & \textbf{28.57}                  & \textbf{32.09 }              & \multicolumn{1}{c|}{{
 \textbf{33.81}}}    & \textbf{14.75 }         \\ \hline
Self-RAG                                                      & 22.41               & 12.09                  & 18.27               & \multicolumn{1}{c|}{21.63}          & 38.33          \\
\begin{tabular}[c]{@{}c@{}}Self-RAG w/ ConTRe\end{tabular} & \textbf{37.58}               & \textbf{17.8   }                & \textbf{23.25}               & \multicolumn{1}{c|}{{\textbf{26.63}}} & { \textbf{34.73}}    \\ \hline
\end{tabular}%
}
\caption{Versatility of ConTRe (Retrieval only)}
\label{tab:ConTRe_performance}
\end{table}
\subsubsection{Versatility of ConTRe}
Our retrieval-only module, ConTRe, is designed to be compatible with any sophisticated generation module. Among the baseline approaches evaluated, only the DSP and Self-RAG frameworks employ distinct generation strategies different from ConTReGen. Specifically, DSP integrates Chain-of-Thought prompting into its generation process, while Self-RAG enhances its fine-tuned large language model (LLM) with additional reflection tokens to improve text generation quality. We integrated ConTRe with both the DSP and Self-RAG generation modules and reported results in Table~\ref{tab:ConTRe_performance} on ASQA. This integration of ConTRe significantly enhances performance in both cases, surpassing their respective default retrieval strategies.
\subsection{Computation Overhead} Since the LLM is the most computationally intensive component, we measured the average number of LLM calls made by each method on the ODSUM-Story dataset, as shown in Table~\ref{tab:llm_calls}. Although ConTReGen requires more LLM or retrieval calls, it uses a smaller LLM (Phi-3, 3.8B) and still significantly outperforms all baselines with larger models (LLAMA3, 8B), achieving a +16.72 improvement in recall, as detailed in Table~\ref{tab:retrieval_varying_llms}. Exploring the trade-offs between computational cost and task performance could be a promising direction for future work.
\begin{table}[]
\centering
\resizebox{\columnwidth}{!}{%
\begin{tabular}{|c|c|c|c|c|c|c|}
\hline
RetGen & IterRetGen & SelfAsk & SearChain & DSP & SelfRAG & ConTReGen \\ \hline
1      & 5          & 6       & 19.56     & 12  & 22      & 44.52     \\ \hline
\end{tabular}%
}
\caption{Number of LLM call}
\label{tab:llm_calls}
\end{table}
\section{Related Works}
Recently, considerable research has focused on Retrieval-Augmented Generation (RAG) across various NLP tasks~\cite{gao2023retrieval}. Existing approaches to single-time retrieval augmented generation typically involved retrieving knowledge directly from the input itself~\cite{izacard2020leveraging,lewis2020retrieval,petroni2020kilt}, using expanded queries~\cite{chuang2023expand}, or rewritten queries~\cite{ma2023query}. However, as these methods often struggle to retrieve all relevant passages in one go, recent developments have shifted towards iterative retrieval approaches. These involve LLMs actively interacting with the retrieval process by formulating contextually rich subsequent queries, which can include utilizing full or partial responses~\cite{shao2023enhancing,trivedi2022interleaving,asai2023self,jiang-etal-2023-active}, or generating new queries~\cite{press2022measuring,xu2024search,yao2022react}. Additionally, adaptive retrieval-augmented approaches have emerged, which decide when to retrieve based on token probability~\cite{jiang-etal-2023-active}, by generating explicit retrieval token~\cite{asai2023retrieval}, or by leveraging the LLM’s self-knowledge. No iterative approaches leverage hierarchical exploration of diverse facets of input queries in-depth. A recent paper MEMWALKER~\cite{chen2023walking} constructs a tree structure from a given long text irrespective of queries. The model traverses the structure upon receiving an input query and generates the response, whereas our ConTReGen constructs a query-focused tree by retrieving the relevant information from a corpus. It focuses on understanding long text while ConTReGen focuses on both retrieving relevant contexts and leveraging them to generate the response.  Similar to MEMWALKER~\cite{chen2023walking}, another recent paper~\cite{sarthi2024raptor} utilizes a bottom-up method to build a hierarchical tree by clustering and summarizing text chunks, transforming a flat corpus into a multi-layered one for varied-level information retrieval. However, this approach may struggle with very large retrieval corpora and highly diversified queries because the clustering and summarization processes are conducted independently of the queries. In contrast, ConTReGen employs a top-down strategy, systematically branching out from the main query into sub-queries that explore different facets and uses query-focused bottom-up summarization, enhancing relevance and effectiveness in information summarization without altering the retrieval corpus.

\section{Conclusion}
In this paper, we introduced ConTReGen, an innovative framework designed to enhance the capabilities of open-domain long-form text generation through a novel context-driven, tree-structured retrieval approach. ConTReGen organizes the retrieval process hierarchically, allowing for an in-depth exploration of various facets of input queries, and integrates a systematic bottom-up synthesis. This approach addresses the prevalent challenges in the field by ensuring comprehensive coverage and coherent integration of multi-faceted information. Additionally, our retrieval-only ConTRe module is model-agnostic and can be seamlessly adapted to various advanced generation techniques. Our extensive experiments across multiple datasets have demonstrated that ConTReGen significantly outperforms existing state-of-the-art retrieval-augmented generation models. These results not only validate the effectiveness of our approach but also underline its potential to be a versatile tool in enhancing the generation of contextually rich, accurate long-form content. Moreover, the introduction of the ODSUM-WikiHow dataset is particularly significant for furthering research in open-domain summarization.

\section*{Acknowledgments}
We thank the anonymous reviewers for their valuable comments and suggestions. This material is based upon work supported by the National Science 
Foundation IIS 16-19302 and IIS 16-33755, Zhejiang University ZJU 
Research 083650, Futurewei Technologies HF2017060011 and 094013, 
IBM-Illinois Center for Cognitive Computing Systems Research (C3SR) and 
IBM-Illinois Discovery Accelerator Institute (IIDAI), grants from eBay 
and Microsoft Azure, UIUC OVCR CCIL Planning Grant 434S34, UIUC CSBS 
Small Grant 434C8U, and UIUC New Frontiers Initiative. Any opinions, 
findings, conclusions, or recommendations expressed in this publication 
are those of the author(s) and do not necessarily reflect the views of 
the funding agencies.

\section*{Limitation}
While ConTReGen has shown promising results in open-domain long-form text generation, its effectiveness is contingent upon the quality of the information retrieved. The model is susceptible to incorporating noisy or irrelevant documents, which can degrade the quality of the generated content. NLI contradiction scores of ConTReGen also highlight the need for improvement in managing irrelevant contexts.  To mitigate this issue, explicitly trained generation models, such as recent works~\cite{yoran2023making,asai2023self} could be employed to enhance the robustness of the generation process against irrelevant contexts. Developing advanced generation techniques tailored for ConTReGen represents a promising direction for future research.

Moreover, the planning agent in ConTReGen, which is crucial for generating and organizing sub-queries, could benefit from explicit training. This training would enable the agent to adaptively generate plans and autonomously determine the necessary depth of exploration for each aspect of a query. Currently, ConTReGen may generate vague or non-informative sub-queries, and its two-step verification process relies heavily on the LLM’s reasoning and existing knowledge base. Despite its impressive retrieval capabilities, there is significant potential for enhancing the system by implementing a more sophisticated model. Such a model would not only verify but also generate contextually rich, retrieval-effective sub-queries, thereby substantially improving both the accuracy and relevance of the retrieval process.

Furthermore, as this work focuses on long-form text generation, it would be an interesting future direction to explore tree-structured retrieval in single factoid answer scenarios.

\bibliography{anthology}

\appendix

\section{Appendix}
\label{sec:appendix}
\begin{table*}[thbp]
\centering
\resizebox{\textwidth}{!}{%
\begin{tabular}{|c|cccc|cccc|}
\hline
\multirow{2}{*}{}           & \multicolumn{4}{c|}{Topk = 3, Iterations = 10 | LLAMA3-8B as LLM}                                                                                                           & \multicolumn{4}{c|}{Topk = 5, Iterations = 10 | Phi-3-Mini-128K as LLM}                                                                                                           \\ \cline{2-9} 
                            & \multicolumn{1}{c|}{{R-L}}   & \multicolumn{1}{c|}{{BERTScore}} & \multicolumn{1}{c|}{{NLI Entailment}} & {Retrieval Recall} & \multicolumn{1}{c|}{{R-L}}   & \multicolumn{1}{c|}{{BERTScore}} & \multicolumn{1}{c|}{{NLI Entailment}} & {Retrieval Recall} \\ \hline
RetGen                      & \multicolumn{1}{c|}{15.19}          & \multicolumn{1}{c|}{48.19}              & \multicolumn{1}{c|}{10.75}                   & 16.14                     & \multicolumn{1}{c|}{15.39}          & \multicolumn{1}{c|}{51.16}              & \multicolumn{1}{c|}{10.82}                   & 21.69                     \\ 
IterRetGen                  & \multicolumn{1}{c|}{13.12}          & \multicolumn{1}{c|}{48.94}              & \multicolumn{1}{c|}{14.82}                   & 29.03                     & \multicolumn{1}{c|}{14.66}          & \multicolumn{1}{c|}{50.86}              & \multicolumn{1}{c|}{14.85}                   & 37.95                     \\ 
Self-Ask                    & \multicolumn{1}{c|}{16.86}          & \multicolumn{1}{c|}{51.13}              & \multicolumn{1}{c|}{15.14}                   & 37.65                     & \multicolumn{1}{c|}{16.46}          & \multicolumn{1}{c|}{51.32}              & \multicolumn{1}{c|}{15.73}                   & 39.13                     \\ 
DSP      & \multicolumn{1}{c|}{17.82}          & \multicolumn{1}{c|}{52.06}              & \multicolumn{1}{c|}{12.92}                   & 28.67                     & \multicolumn{1}{c|}{16.85}          & \multicolumn{1}{c|}{50.36}              & \multicolumn{1}{c|}{10.57}                   & 32.93                     \\ 
Self-RAG                    & \multicolumn{1}{c|}{18.96}          & \multicolumn{1}{c|}{47.96}              & \multicolumn{1}{c|}{9.37}                    & 23.24                     & \multicolumn{1}{c|}{17.78}          & \multicolumn{1}{c|}{47.83}              & \multicolumn{1}{c|}{11.34}                   & 29.28                     \\ 
\textit{\textbf{ConTReGen}} & \multicolumn{1}{c|}{\textbf{19.33}} & \multicolumn{1}{c|}{\textbf{54.01}}     & \multicolumn{1}{c|}{\textbf{21.35}}          & \textbf{57.69}            & \multicolumn{1}{c|}{\textbf{18.36}} & \multicolumn{1}{c|}{\textbf{53.42}}     & \multicolumn{1}{c|}{\textbf{18.15}}          & \textbf{53.51}            \\ \hline
\end{tabular}%
}
\caption{Performance comparison by varying the number of iterations and Top k on ODSUM-Story dataset}
\label{tab:varying_iterations}
\end{table*}
\begin{table*}[]
\centering
\resizebox{\textwidth}{!}{%
\begin{tabular}{|c|c|c|c|c|c|c|c|}
\hline
RetGen & IterRetGen & SelfAsk & SearChain & DSP   & SelfRAG & ConTReGen (Phi-3) & ConTReGen (LLAMA-3) \\ \hline
16.14  & 26.98      & 36.79   & 15.24     & 27.61 & 21.14   & 53.51             & 57.59               \\ \hline
\end{tabular}%
}
\caption{Retrieval Performance of ConTReGen with different LLMs}
\label{tab:retrieval_varying_llms}
\end{table*}
\subsection{Experimental results by varying hyperparameters}
We analyzed the performance of several iterative retrieval techniques per iteration on ODSUM-Story and reported the results in Figure~\ref{fig:retrieverl_per_iterations}, which highlights that they are likely to reach a plateau of retrieval recall very quickly. Additionally, Table~\ref{tab:varying_iterations} shows the performance comparison across baselines after 10 iterations, while keeping the setting of ConTReGen as the number of subquestions in a plan is 5, and the depth of tree is 2.

\subsection{Retrieval Performance by varying LLMs}
Considering resource-constrained scenarios, we conducted further experiments using the lightweight Phi-3-Mini-4K LLM in our retrieval module. The recall performance on the ODSUM-Story dataset demonstrates that our Phi-3 retrieval significantly outperforms baseline approaches that use the LLAMA-3-8B model as shown in Table~\ref{tab:retrieval_varying_llms}


\subsection{Error Analysis}
We have conducted a detailed error analysis to identify scenarios where the input query might not be sufficiently informative to retrieve relevant reference passages from the corpus. Specifically, we have found 10 instances within the ODSUM-Story dataset where the input query, when used as the retriever’s query, did not yield any evidence passages in the Top 32 results. In these challenging cases:
\begin{itemize}
    \item All other baselines: Failed to retrieve any evidence passages in every case.
    \item \textbf{ConTReGen}: Successfully retrieved at least one evidence passage in 3 out of the 10 cases.
\end{itemize}
To further illustrate the robustness and reliability of ConTReGen, consider the following case where the original query is too vague considering the corpus:

\textbf{Original Query}: “What is the storyline of PRISON PLANET?”

\textbf{Result:} All approaches including retriever (topk=32) and ConTReGen failed to retrieve any evidence passages from the corpus.

Upon modifying the original query to include the author's name:
\textbf{Modified Query:} “What is the storyline of PRISON PLANET by BOB TUCKER?”

\textbf{Result:}
\begin{itemize}
    \item All baselines failed, except for DSP and Retriever (topk=32) retrieved just 1 evidence passage out of 9.
    \item \textbf{ConTReGen}: Retrieved 6 evidence passages out of 9.
\end{itemize}

The identified error cases underline a crucial aspect of query informativeness. In scenarios where the input query lacks informativeness, all approaches including ConTReGen face challenges in retrieving relevant passages. However, ConTReGen shows a notable improvement over all baselines when the query is slightly modified to include additional context, such as the author's name. This suggests that while ConTReGen is more robust and reliable, its performance can still be influenced by the informativeness of the input query.

\end{document}